\documentstyle[pra,aps,twocolumn,floats,epsf]{revtex}

\begin{document}

\draft

\title{Correlated scattering of photons from a trapped Bose-Einstein
condensate}

\author{Hiroki Saito and Masahito Ueda}
\address{Department of Physical Electronics, Hiroshima University,
Higashi-Hiroshima 739-8527, Japan \\
and CREST, Japan Science and Technology Corporation (JST),
Higashi-Hiroshima 739-8527, Japan}


\maketitle

\begin{abstract}
The second-order coherence of photons scattered from a trapped
Bose-Einstein condensate is found to be enhanced for the scattering angles
that are either the same or symmetrical with respect to the direction of
laser propagation.
The enhancement occurs periodically with respect to the time interval
between photons, and becomes less and less pronounced with increasing the
strength of the interaction between atoms.
\end{abstract}

\narrowtext

\section{Introduction}

Realization of Bose-Einstein condensation (BEC) in trapped atomic
gases~\cite{Anderson,Bradley,Davis,Fried} has created an interdisciplinary 
subfield of physics called coherent matter wave.
Developments of optical imaging techniques such as absorption imaging and
a nondestructive {\it in situ} method~\cite{Andrews96} have enabled us to
observe interference between matter waves~\cite{Andrews97} and dynamics
of BEC~\cite{AndrewsPRL,Hall}.
In a remarkable recent experiment, Hau {\it et al.}~\cite{Hau}
demonstrated that it is possible to make photons propagate through BEC at
a speed of 17 m/s, which suggests that the index of refraction of BEC is
14 orders of magnitude greater than that of a glass fiber.
In the present paper we predict yet another unique feature of BEC
concerning correlated photon scattering.

Various aspects of the optical response of BEC have been
predicted~\cite{Svistunov,Politzer,Lewen93,Java94,Lewen94,You94,Java95,You95,Morice,JavaRuo95,You96}:
a broad resonance~\cite{Java94,You94}, an extra structure in the
spectrum~\cite{Java95,JavaRuo95}, response to intense short
pulses~\cite{Lewen93,You95}, non-Lorentzian line shapes~\cite{You94,You96}, 
and a refractive index of a homogeneous gas~\cite{Morice}.
In the above work, kinetic motion of atoms is either ignored or not
explicitly taken into account because its time scale is usually much
slower than that of optical processes.
However, optical signatures of BEC are expected to appear also in the time
scale of kinetic motion, since the characteristic feature of BEC is the
degeneracy of the kinetic degrees of freedom as well as the atomic
internal degree of freedom.

The aim of the present paper is to study correlations of photons scattered
from trapped BEC that is irradiated by a weak far-off-resonant laser
field, and to show that they exhibit some characteristic features
associated with kinetic motion of the atoms in the trap.
By scattering a photon, the condensate suffers a recoil, which affects the
scattering of the subsequent photons.
As a consequence a correlation arises between a pair of photons with
respect to their scattering angles and time intervals.
We show that the presence of the condensate enhances the probability of a
pair of photons to be scattered into either the same direction or
symmetrical directions about laser propagation, and the enhancement occurs 
periodically with respect to the time interval, where the period is set
by the frequency of the trap.

The interaction between atoms cannot be ignored, for the profile of the
condensate is significantly altered even for a few hundred
atoms~\cite{Ruprecht,Dalfovo96}.
While in Ref.~\cite{You96} effects of the atom-atom interaction on the
photon scattering are taken into account only through the atomic density
profile, the dynamical effect of the interaction on the optical response
has not been discussed.
We take into account the atom-atom interaction using the Bogoliubov
approximation~\cite{Bogoliubov}, and show that the correlations between
the scattered photons are deteriorated by the interaction.

This paper is organized as follows.
Section \ref{s:source} formulates the problem and derives an expression of
the electric-field operator far from BEC irradiated by a laser field.
Section \ref{s:noninteraction} studies coherence properties of scattered
photons.
Section~\ref{s:interaction} discusses effects of the atom-atom
interaction on the scattering of photons, and Sec.~\ref{s:conclusion}
concludes this paper.

\section{Formulation of the Problem}
\label{s:source}

We consider a situation in which trapped atoms are irradiated by a laser
field which is sufficiently detuned from the resonant frequency of the
atoms so that each photon is not scattered by the atoms more than once.
Photons are scattered from the atoms and detected by photon counters.
Quantum-statistical properties of the scattered photons can be
studied through the electric-field operator $\hat {\bf E}({\bf r}, t)$ in
the Heisenberg representation.
We derive here an expression of $\hat {\bf E}({\bf r}, t)$ at the position
far away from the trapped atoms compared with the spread of their wave
functions.

The Hamiltonian for the photon field is given by
\begin{equation}
\hat H_{\rm f} = \sum_{{\bf k}\sigma} \hbar \omega_{\bf k} \hat a_{{\bf k}
\sigma}^\dagger \hat a_{{\bf k} \sigma},
\end{equation}
where $\hat a_{{\bf k} \sigma}^\dagger$ and $\hat a_{{\bf k} \sigma}$ are
the creation and annihilation operators of a photon with wave vector $\bf
k$ and polarization $\sigma$.
The electric-field operator is expanded in terms of the creation and
annihilation operators as
\begin{equation}
\hat{\bf E}({\bf r}) = i \sum_{{\bf k} \sigma} \sqrt{\frac{\hbar
\omega_k}{2 \varepsilon_0 V}} \left( \bbox{\epsilon}_{{\bf k} \sigma} e^{i
{\bf k} \cdot {\bf r}} \hat a_{{\bf k} \sigma} - \bbox{\epsilon}_{{\bf k}
\sigma}^* e^{-i {\bf k} \cdot {\bf r}} \hat a_{{\bf k} \sigma}^\dagger
\right),
\end{equation}
where $\bbox{\epsilon}_{{\bf k} \sigma}$ is the polarization vector of
photons.
The internal states of two-level atoms are described by the Hamiltonian
\begin{equation} \label{H_AI}
\hat H_{\rm ai} = \sum_{j = 1}^N \frac{\hbar \omega_A}{2} \hat
\sigma_{jz},
\end{equation}
where $N$ is the number of atoms, $\hbar \omega_A$ is the transition
energy of the internal levels, and the operator $\hat \sigma_{jz}$ is the
Pauli operator which takes on the eigenvalue 1 (or $-1$) when the $j$th
atom is in the excited (or ground) internal level.
The Hamiltonian corresponding to the kinetic degrees of freedom of the
atoms is given by
\begin{equation} \label{H_AK}
\hat H_{\rm ak} = \sum_{j = 1}^N \left[ -\frac{\hbar^2}{2m} \nabla_j^2 +
V_t({\bf R}_j) \right] + \sum_{j \neq j'}^N \frac{2 \pi \hbar^2 a}{m}
\delta({\bf R}_j - {\bf R}_{j'}),
\end{equation}
where $m$ is the mass of the atom and $V_t({\bf R})$ is the trap
potential.
The last term in Eq.~(\ref{H_AK}) describes the Fermi's contact type of
interaction between hard-core atoms, which is characterized by the s-wave
scattering length $a$.

We assume the electric-dipole interaction between the atoms and the photon 
field which is described by the Hamiltonian
\begin{equation} \label{H_ED}
\hat H_{\rm ed} = -\sum_{j = 1}^N \hat{\bf D}_j \cdot \hat{\bf E}({\bf
R}_j).
\end{equation}
The dipole operator denoted as $\hat{\bf D}_j$ has the form
\begin{equation}
\hat{\bf D}_j \equiv {\bf d} \hat \sigma_{j+} + {\bf d}^* \hat
\sigma_{j-},
\end{equation}
where $\bf d$ is the electric-dipole matrix element, and $\hat
\sigma_{j+}$ and $\hat \sigma_{j-}$ are the raising and lowering operators
of the internal state of the $j$th atom.

The Hamiltonian for the atom-laser interaction is given by
\begin{eqnarray}
\hat H_L & = & \sum_{j = 1}^N \Bigl[ {\cal E}_L (\bbox{\epsilon}_L \cdot
{\bf  d}) e^{-i (\omega_L t - {\bf k}_L \cdot {\bf R}_j)} \hat \sigma_{j+} 
\nonumber \\
& & + {\cal E}_L^* (\bbox{\epsilon}_L \cdot {\bf d})^* e^{i (\omega_L t -
{\bf k}_L \cdot {\bf R}_j)} \hat \sigma_{j-} \Bigr],
\end{eqnarray}
where ${\cal E}_L$, $\bbox{\epsilon}_L$, $\omega_L$, and ${\bf k}_L$
denote the amplitude, the polarization vector, the frequency, and the wave 
vector of the laser field, respectively.
Since the optical processes of the internal states of the atoms are much
faster than those of the kinetic motion, we may adiabatically eliminate the
dynamics of the internal state.
When the detuning $\delta \equiv \omega_L - \omega_A$ is much larger than
both the atom-laser field coupling and the collective linewidth $\Gamma$
of the atoms~\cite{Java94,You94}, i.e.,
\begin{equation}
|\delta| \gg \frac{|{\cal E}_L {\bf d}|}{\hbar}, \;\;\; |\delta| \gg
\Gamma,
\end{equation}
the probability that the atoms are in the excited state $|e \; \rangle$ is
much smaller than that in the ground state $|g \; \rangle$.
In this case, the atomic internal state may be assumed to be in the ground
state of the Hamiltonian $\hat H_{\rm ai} + \hat H_{L}$, which is given,
up to the first order in $|{\cal E}_L {\bf d}| / \hbar \delta$, by
\begin{equation} \label{groundstate}
\prod_{j = 1}^N \left[ e^{i(\omega_L t - {\bf k}_L \cdot {\bf R}_j) / 2}
|g \; \rangle_j + \frac{|{\cal E}_L {\bf d}|}{\hbar \delta} e^{-i(\omega_L
t - {\bf k}_L \cdot {\bf R}_j) / 2} |e \; \rangle_j \right].
\end{equation}
Taking the expectation value of $\hat H_{\rm ed}$ in Eq.~(\ref{H_ED})
with respect to the atomic internal state (\ref{groundstate}), we obtain
the effective Hamiltonian describing the interaction between photons and
kinetic motion of the atoms via the atomic internal levels as
\begin{eqnarray}
\hat H_{\rm ed}' & = & \sum_{j = 1}^N \frac{|{\cal E}_L {\bf d}|}{\hbar
\delta} 
\Bigl[ e^{i(\omega_L t - {\bf k}_L \cdot {\bf R}_j)} {\bf d} \nonumber \\
& & +
e^{-i(\omega_L t - {\bf k}_L \cdot {\bf R}_j)} {\bf d}^* \Bigr] \cdot
\hat{\bf E}({\bf R}_j).
\end{eqnarray}
In the following discussions we will employ the effective Hamiltonian
$\hat H_{\rm eff} = \hat H_{\rm f} + \hat H_{\rm ak} + \hat H_{\rm ed}'$
to describe the dynamics of photons and kinetic motion of the atoms.

The equation of motion for the electric-field operator in the Heisenberg
representation is obtained from the Hamiltonian $\hat H_{\rm eff}$ as
\begin{equation} \label{eomE}
\left( \nabla^2 - \frac{\partial^2}{c^2 \partial t^2} \right)
\hat{\bf E}({\bf r}, t) = \frac{i}{\hbar} [ \hat H_{\rm ed}'(t), \nabla^2
\hat{\bf A}({\bf r}, t) ],
\end{equation}
where the vector potential is given by
\begin{eqnarray}
\hat{\bf A}({\bf r}, t) & = & \sum_{{\bf k} \sigma} \sqrt{\frac{\hbar}{2
\varepsilon_0 \omega_k V}} \Bigl[ e^{i {\bf k} \cdot {\bf r}}
\bbox{\epsilon}_{{\bf k} \sigma} \hat a_{{\bf k} \sigma}(t) \nonumber \\
& & + e^{-i {\bf
k} \cdot {\bf r}} \bbox{\epsilon}_{{\bf k} \sigma}^* \hat a_{{\bf k}
\sigma}^\dagger(t) \Bigr].
\end{eqnarray}
The equation of motion (\ref{eomE}) can be solved, giving
\begin{equation}
\hat{\bf E}({\bf r}, t) = \hat{\bf E}_{\rm vac}({\bf r}, t) + \hat{\bf
E}_{\rm s}({\bf r}, t),
\end{equation}
where $\hat{\bf E}_{\rm vac}({\bf r}, t) = e^{\frac{i}{\hbar} \hat H_{\rm
f}  t} \hat{\bf E}({\bf r}) e^{-\frac{i}{\hbar} \hat H_{\rm f}  t}$ is the
vacuum field and
\begin{eqnarray} \label{sourcef}
\hat{\bf E}_{\rm s}({\bf r}, t) & = & -\frac{i}{4 \pi \hbar} \int d{\bf
r}' dt' \frac{[ \hat H_{\rm ed}'(t'), \nabla^2 \hat{\bf A}({\bf r}', t')
]}{|{\bf r} - {\bf r}'|} \nonumber \\
& & \times \delta(t - t' - |{\bf r} - {\bf r}'| / c)
\end{eqnarray}
describes the source field scattered from the atoms.
Evaluating the commutator and carrying out the integral with respect to
$t'$, the positive frequency part of the source field (\ref{sourcef})
becomes
\begin{eqnarray} \label{sf1}
\hat{\bf E}_{\rm s}^{(+)}({\bf r}, t) & = & -\frac{1}{4 \pi \varepsilon_0 V}
\frac{|{\cal E}_L {\bf d}|}{\hbar \delta} \int d{\bf r}' \frac{e^{-i
\omega_L t_r}}{|{\bf r} - {\bf r}'|} \nonumber \\
& & \times \sum_{\bf k} {\bf k} \times ({\bf d}
\times {\bf k}) e^{-i {\bf k} \cdot {\bf r}'} \sum_{j = 1}^N e^{i ({\bf
k}_L + {\bf k}) \cdot {\bf R}_j(t_r)}, \nonumber \\
\end{eqnarray}
where $t_r \equiv t - |{\bf r} - {\bf r}'| / c$ and ${\bf R}_j(t_r) \equiv 
e^{\frac{i}{\hbar} \hat H_{\rm eff} t_r} {\bf R}_j e^{-\frac{i}{\hbar}
\hat H_{\rm eff} t_r}$.
Since $|{\cal E}_L {\bf d}| / \hbar \delta \ll 1$, it is sufficient to
keep only the terms first order in $|{\cal E}_L {\bf d}| / \hbar \delta$,
which amounts to disregarding the multiple scattering and the
dipole-dipole interaction.
Consequently, we can make an approximation
\begin{equation} \label{rapp}
{\bf R}_j(t_r) \simeq e^{\frac{i}{\hbar} \hat H_{\rm ak} t} {\bf R}_j
e^{-\frac{i}{\hbar} \hat H_{\rm ak} t} \equiv {\bf R}_j(t),
\end{equation}
where we ignore the time $|{\bf r} - {\bf r}'| / c$ it takes photons to
propagate from the atoms to the detector as it is much smaller than the
time scale of the kinetic motion of the atoms.
The integrand of the source field (\ref{sf1}) contributes only when ${\bf
r}'$ falls within the spread of the atomic wave function.
Therefore, if we choose the origin of the coordinate system at the center
of the trap, the distance $|{\bf r}|$ from the atom to the detector is
much larger than $|{\bf r}'|$, and hence $t_r \simeq t - (r / c) \left( 1
- {\bf r} \cdot {\bf r}' / r^2 \right)$ and $1 / |{\bf r} - {\bf r}'|
\simeq 1 / r$.
We thus obtain
\begin{eqnarray} \label{sourceE}
\hat {\bf E}_{\rm s}^{(+)}({\bf r}, t) & = & -\frac{1}{4 \pi \varepsilon_0 r V}
\frac{|{\cal E}_L {\bf d}|}{\delta \hbar} e^{-i \omega_L t} \int d{\bf
r}' \sum_{\bf k} {\bf k} \times ({\bf d} \times {\bf k}) \nonumber \\
& & \times e^{ikr} e^{i (k_L
\tilde{\bf r} - {\bf k}_L) \cdot {\bf r}'} \sum_{j = 1}^N e^{i ({\bf k}_L
+ {\bf k}) \cdot {\bf R}_j(t)} \nonumber \\
& = & -\frac{k_L^2}{4 \pi \varepsilon_0 r} \frac{|{\cal E}_L {\bf
d}|}{\delta \hbar} {\bf n} \times ({\bf d} \times {\bf n}) e^{-i \omega_L
t + i k_L r} \nonumber \\
& & \times \sum_{j = 1}^N e^{i ({\bf k}_L - k_L {\bf n}) \cdot {\bf
R}_j(t)} \nonumber \\
& \equiv & {\bf F}({\bf r}, t) \sum_{j = 1}^N e^{i \Delta {\bf k}
\cdot {\bf R}_j(t)},
\end{eqnarray}
where ${\bf n} \equiv {\bf r} / |{\bf r}|$ and $\Delta {\bf k} \equiv {\bf
k}_L - k_L {\bf n}$.
The expression of the source-field operator (\ref{sourceE}) consists of
the factor ${\bf F}({\bf r}, t)$ which depends on the spatial
configuration of the system and the operator that transfers a recoil
momentum $\hbar \Delta {\bf k}$ to the atoms.

\section{First and Second-Order Coherence of photons scattered from a
noninteracting condensate}
\label{s:noninteraction}

In this section we consider an ideal Bose gas, where the atom-atom
interaction is absent.
We write the eigenfunctions of the Hamiltonian $\hat H_{\rm ak}$ as
$f_n({\bf R})$, where $n$ is an index for the eigenstates with $n = 0$
referring to the ground state.

It is convenient to discuss the problem in the second-quantized
formalism.
We expand the atomic field operator as
\begin{equation} \label{psi}
\hat \psi({\bf R}) = \sum_n f_n({\bf R}) \hat b_n,
\end{equation}
where $\hat b_n$ annihilates an atom in the $n$th eigenstate, and
satisfies the Bose commutation relation $[\hat b_n, \hat b_{n'}^\dagger]
= \delta_{nn'}$.
The Hamiltonian $\hat H_{\rm ak}$ can then be rewritten as
\begin{equation}
\hat H_{\rm ak} = \sum_n \hbar \omega_n \hat b_n^\dagger b_n,
\end{equation}
where $\hbar \omega_n$ is the $n$th eigenenergy.
The second-quantized form of the operator $\sum_{j = 1}^N e^{i \Delta {\bf
k} \cdot {\bf R}_j (t)}$ in Eq.~(\ref{sourceE}) is given by
\begin{eqnarray} \label{D}
\hat{\cal D}(\Delta {\bf k}, t) & \equiv & \int d{\bf R} \hat
\psi^\dagger({\bf R}, t) e^{i \Delta {\bf k} \cdot {\bf R}} \hat \psi({\bf 
R}, t) \nonumber \\
& = & \sum_{nn'} \langle n' | e^{i \Delta {\bf k} \cdot {\bf R}} | n
\rangle \hat b_{n'}^\dagger \hat b_n e^{i (\omega_{n'} - \omega_n) t}, 
\end{eqnarray}
where $\hat\psi({\bf r}, t) \equiv e^{\frac{i}{\hbar} \hat H_{\rm ak} t}
\hat\psi({\bf R}) e^{-\frac{i}{\hbar} \hat H_{\rm ak} t}$, and the bracket
$\langle n' | e^{i \Delta {\bf k} \cdot {\bf R}} | n \rangle$ is a
shorthand notation of $\int d{\bf R} f_{n'}^*({\bf R}) e^{i \Delta {\bf k}
\cdot {\bf R}} f_n({\bf R})$.
The source-field operator (\ref{sourceE}) is expressed as $\hat{\bf
E}^{(+)}({\bf r}, t) = {\bf F}({\bf r}, t) \hat{\cal D}(\Delta {\bf k},
t)$.

At zero temperature all $N$ atoms are Bose-Einstein condensed in the
ground-state level of the trap potential.
We assume an isotropic harmonic trap $V_t({\bf R}) = m \omega^2
R^2 / 2$ for simplicity.
The expectation value of the electric field at the position $\bf r$ becomes
\begin{eqnarray} \label{Eavg}
\langle \hat{\bf E}^{(+)}({\bf r}, t) \rangle & = & {\bf F}({\bf r}, t) N
\langle 0 | e^{i \Delta {\bf k} \cdot {\bf R}} | 0 \rangle \nonumber \\
& = & {\bf F}({\bf r}, t) N e^{-|d \Delta {\bf k}|^2 / 4},
\end{eqnarray}
where $d \equiv \left( \hbar / m \omega \right)^{1/2}$ denotes the
characteristic length scale of the trap.
The exponential factor in Eq.~(\ref{Eavg}) is a decreasing function of the 
scattering angle.

\begin{figure}[tb]
\begin{center}
\leavevmode\epsfxsize=86mm \epsfbox{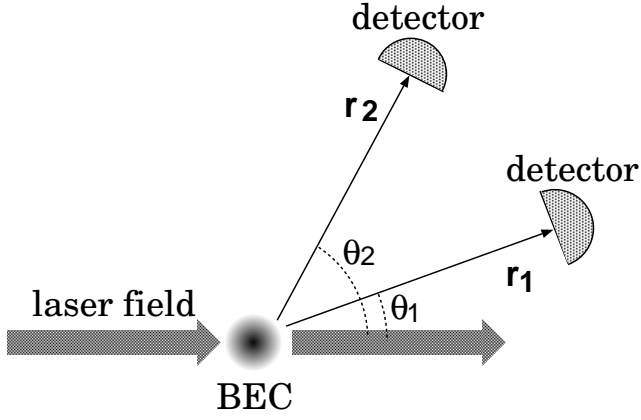}
\end{center}
\caption{
Schematic of a scattering experiment.
The trapped Bose-Einstein condensate (BEC) is irradiated by a weak and
far off-resonant laser.
The scattered photons are detected by the detectors set around the BEC.
The positions of the detectors are denoted as ${\bf r}_i$ and the
scattering angles as $\theta_i$ ($i = 1, 2$).
}
\label{f:setup}
\end{figure}

Suppose that we detect the number of scattered photons in a configuration
schematically illustrated in Fig.~\ref{f:setup}.
The position of the $i$th detector is denoted as ${\bf r}_i$ and the
scattering angle as measured from the direction of laser propagation is
denoted as $\theta_i$.
The number of photons counted by the detector is proportional to the
normally-ordered intensity of the field~\cite{Loudon}
\begin{equation}
I({\bf r}, t) \equiv \langle \hat{\bf E}^{(-)}({\bf r}, t) \cdot \hat{\bf
E}^{(+)}({\bf r}, t) \rangle.
\end{equation}
Substituting Eq.~(\ref{sourceE}) into this gives
\begin{equation} \label{I}
I({\bf r}, t) = |{\bf F}({\bf r}, t)|^2 N \left[ 1 + (N - 1)
e^{-|d \Delta {\bf k}|^2 / 2} \right].
\end{equation}
The first term on the right-hand side is proportional to the number of
atoms $N$ and the angular dependence lies solely in $|{\bf F}({\bf r},
t)|^2$.
The second term is proportional to $N(N - 1)$ and contains a
forward-scattering factor of $e^{-|d \Delta {\bf k}|^2 / 2}$.
In fact, in experiments of Ref.~\cite{Andrews96}, a forward scattering
that satisfies $|d \Delta {\bf k}| \lesssim 1$ was observed at the onset
of BEC.
The degree of first-order coherence is defined by~\cite{Loudon}
\begin{equation}
g^{(1)}({\bf r}_1, t_1; {\bf r}_2, t_2) \equiv \frac{\langle \hat{\bf
E}^{(-)}({\bf r}_1, t_1) \cdot \hat{\bf E}^{(+)}({\bf r}_2, t_2)
\rangle}{\sqrt{I({\bf r}_1, t_1) I({\bf r}_2, t_2)}}.
\end{equation}
Substituting Eq.~(\ref{sourceE}) into this gives
\begin{eqnarray} \label{g1}
& & |g^{(1)}({\bf r}_1, t_1; {\bf r}_2, t_2)| \nonumber \\
& = & \frac{e^{-|d \Delta {\bf
k}_1|^2 / 4 - |d \Delta {\bf k}_2|^2 / 4} \left| e^{d^2 \Delta {\bf k}_1
\cdot \Delta {\bf k}_2 e^{-i \omega (t_1 - t_2)} / 2} + N - 1
\right|}{\left [ 1 + (N - 1) e^{-|d \Delta {\bf k}_1|^2 / 2}
\right]^{\frac{1}{2}} \left[ 1 + (N - 1) e^{-|d \Delta {\bf k}_2|^2 / 2}
\right]^{\frac{1}{2}}}, \nonumber \\
\end{eqnarray}
where $\Delta {\bf k}_i \equiv {\bf k}_L - k_L {\bf r}_i / |{\bf r}_i|$
($i = 1$ and $2$).
When the number of atoms is sufficiently large to satisfy $N e^{-|d \Delta 
{\bf k}_i|^2 / 2} \gg 1$ for $i = 1$ and $2$, the degree of the
first-order coherence becomes $|g^{(1)}({\bf r}_1, t_1; {\bf r}_2, t_2)|
\simeq 1$.
The scattered photon field is therefore first-order coherent
when the number of atoms in the condensate is sufficiently large.
On the other hand, when $N e^{-|d \Delta {\bf k}_i|^2 / 2} \ll 1$ for both 
$i = 1$ and $2$, Eq.~(\ref{g1}) reduces to 
\begin{equation} \label{g1r}
|g^{(1)}({\bf r}_1, t_1; {\bf r}_2, t_2)| \simeq \exp\left(-\frac{d^2}{4}
|\Delta {\bf k}_1 e^{-i \omega (t_1 - t_2)} - \Delta {\bf k}_2 |^2
\right).
\end{equation}
When $t_1 = t_2$, the coherence (\ref{g1r}) is lost exponentially with
increasing the difference in the scattering angles $|\Delta {\bf k}_1 -
\Delta {\bf k}_2|$.
This is because two photon paths corresponding to atoms kicked in
different directions do not interfere with each other.
The coherence is periodic with respect to the time interval $t_1 - t_2$
with the period of $2\pi \omega^{-1}$.
We note that the same expression as Eq.~(\ref{g1r}) can be obtained by
substituting $N = 1$ in Eq.~(\ref{g1}), and therefore the correlation
described above cannot be ascribed to BEC.

\begin{figure}[tb]
\begin{center}
\leavevmode\epsfxsize=86mm \epsfbox{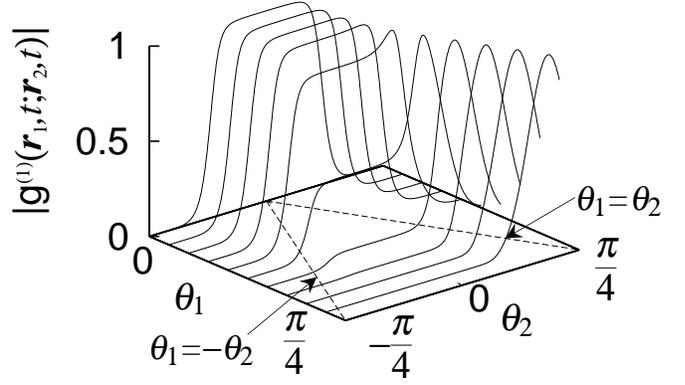}
\end{center}
\caption{
The angular dependence of the degree of first-order coherence
$g^{(1)}({\bf r}_1, t; {\bf r}_2, t)$ of photons scattered from $10^3$
BEC atoms.
The wavelength of the laser field $2 \pi / k_L = 589 {\rm nm}$ (the 3s-3p
transition of ${}^{23}{\rm Na}$) and  the characteristic length of the
trap $d = 1 \mu{\rm m}$ are assumed.
The wave vectors ${\bf k}_1$ and ${\bf k}_2$ are assumed to be in the same 
plane, and $\theta_1$ and $\theta_2$ denote their angles as measured from
${\bf k}_L$.
}
\label{f:g1}
\end{figure}

Figure~\ref{f:g1} shows the degree of first-order coherence for $t_1 =
t_2$, $|g^{(1)}({\bf r}_1, t; {\bf r}_2, t)|$, where $N = 10^3$, $d = 1
\mu{\rm m}$, and $2 \pi / k_L = 589 {\rm nm}$ (the 3s-3p transition of
${}^{23}{\rm Na}$) are assumed.
The wave vector ${\bf k}_1$ and ${\bf k}_2$ are assumed to be in the same
plane, and $\theta_1$ and $\theta_2$ denote the angles of ${\bf k}_1$ and
${\bf k}_2$ as measured from ${\bf k}_L$ (see Fig.~\ref{f:setup}).
One can see that $g^{(1)} \simeq 1$ for $|d \Delta {\bf k}| \lesssim 1$,
and therefore, the photons scattered in the forward directions have
first-order coherence.
There are peaks along $\theta_1 = \theta_2$, which can be seen from
Eq.~(\ref{g1r}).

The degree of second-order coherence~\cite{Loudon} is defined by
\begin{eqnarray} \label{g2}
& & g^{(2)}({\bf r}_1, t_1; {\bf r}_2, t_2) \equiv \nonumber \\
& & \frac{{\displaystyle
\sum_{\alpha, \beta = x, y, z}} \langle \hat E_\alpha^{(-)}({\bf r}_1,
t_1) \hat E_\beta^{(-)}({\bf r}_2, t_2) \hat E_\beta^{(+)}({\bf r}_2, t_2)
\hat E_\alpha^{(+)}({\bf r}_1, t_1) \rangle}{I({\bf r}_1, t_1) I({\bf
r}_2, t_2)}. \nonumber \\
\end{eqnarray}
Substituting Eq.~(\ref{sourceE}) into this yields
\begin{eqnarray} \label{g2_2}
& & g^{(2)}({\bf r}_1, t_1; {\bf r}_2, t_2) = \nonumber \\
& & \Bigl\{ N(N - 1)(N - 2)(N -
3) e^{-|d \Delta {\bf k}_1|^2 / 2 - |d \Delta {\bf k}_2|^2 / 2} \nonumber
\\
& & + N(N - 1)(N - 2) \Bigl[ 2 e^{-|d \Delta {\bf k}_1|^2 / 2 - |d \Delta
{\bf k}_2|^2 / 2} \nonumber \\
& & \times {\rm Re} \Bigl( e^{-d^2 \Delta {\bf k}_1 \cdot \Delta
{\bf k}_2 e^{i \omega (t_2 - t_1)} / 2} + e^{d^2 \Delta {\bf k}_1 \cdot
\Delta {\bf k}_2 e^{-i \omega (t_2 - t_1)} / 2} \Bigr) \nonumber \\
& & + e^{-|d \Delta {\bf k}_1|^2 / 2} + e^{-|d \Delta {\bf k}_2|^2 / 2}
\Bigr] \nonumber \\
& & + N(N - 1) \Bigl[ 1 + e^{-d^2 |\Delta {\bf k}_1 / 2 - \Delta {\bf k}_2
e^{i \omega (t_2 - t_1)}|^2} \nonumber \\
& & + e^{-d^2 |\Delta {\bf k}_1 + \Delta {\bf
k}_2 e^{i \omega (t_2 - t_1)}|^2 / 2}
+ 2 e^{-|d \Delta {\bf k}_1|^2 / 2} \nonumber \\
& & + 2 e^{-|d \Delta {\bf k}_2|^2 / 2}
\cos\bigl( d^2 \Delta {\bf k}_1 \cdot \Delta {\bf k}_2 \sin \omega (t_2
- t_1) \bigr) \Bigr] + N \Bigr\} \nonumber \\
& & \Bigr/ \left\{ N^2 \left[ 1 + (N - 1) e^{-|d \Delta {\bf k}_1|^2 / 2}
\right] \left[ 1 + (N - 1) e^{-|d \Delta {\bf k}_2|^2 / 2} \right]
\right\}. \nonumber \\
\end{eqnarray}
If the number of atoms is sufficiently large to satisfy $N e^{-|d \Delta
{\bf k}_i|^2 / 2} \gg 1$ for both $i = 1$ and $2$, the degree of
second-order coherence becomes $g^{(2)}({\bf r}_1, t_1; {\bf r}_2, t_2)
\simeq 1$ as well as $|g^{(1)}({\bf r}_1, t_1; {\bf r}_2, t_2)| \simeq 1$.
In general, $r$th-order coherence~\cite{Loudon} becomes unity for all $r$, 
and therefore, in this large $N$ limit the scattered photons are coherent
in all orders.
On the other hand, if $N e^{-|d \Delta {\bf k}_i|^2 / 2} \ll 1$ for both
$i = 1$ and $2$, Eq.~(\ref{g2_2}) reduces to
\begin{eqnarray} \label{g2r}
g^{(2)}({\bf r}_1, t_1; {\bf r}_2, t_2) & \simeq & 1 + e^{-d^2 |\Delta {\bf
k}_1 - \Delta {\bf k}_2 e^{i \omega (t_2 - t_1)}|^2 / 2} \nonumber \\
& & + e^{-d^2 |\Delta
{\bf k}_1 + \Delta {\bf k}_2 e^{i \omega (t_2 - t_1)}|^2 / 2}.
\end{eqnarray}
When $t_1 = t_2$, Eq.~(\ref{g2r}) becomes $g^{(2)} \simeq 2$ for either
$\Delta {\bf k}_1 \simeq \Delta {\bf k}_2$ or $\Delta {\bf k}_1 \simeq
-\Delta {\bf k}_2$.
This means that a pair of photons tends to be scattered either in
the same direction or in the symmetrical directions with respect to
laser propagation.
These correlation in scattered photons can be interpreted as follows.
In the case of $\Delta {\bf k}_1 \simeq \Delta {\bf k}_2$, the first
photon kicks an atom from the condensate to the state $e^{i \Delta {\bf
k}_1 \cdot {\bf R}} f_0({\bf R})$, and so does the subsequent photon,
where $f_0({\bf R})$ is the ground state of the trap potential.
Consequently, the Bose-enhancement factor of this process is
\begin{equation}
\langle (N - 2)_0, 2_{\Delta {\bf k}_1} | \hat b_{\Delta {\bf
k}_2}^\dagger \hat b_0 \hat b_{\Delta {\bf k}_1}^\dagger \hat b_0 | N_0,
0_{\Delta {\bf k}_1} \rangle \simeq \sqrt{2 N(N - 1)},
\end{equation}
where $\hat b_{\bf k}^\dagger \equiv \sum_n \langle n | e^{i {\bf k} \cdot
{\bf R}} | 0 \rangle \hat b_n^\dagger$ and $|m_{\bf k} \rangle \equiv \hat 
b_{\bf k}^{\dagger m} / \sqrt{m!} |0 \rangle$.
Therefore the probability that this process occurs is enhanced by a factor
of two.
In the case of $\Delta {\bf k}_1 \simeq -\Delta {\bf k}_2$, there is a
process in which the subsequent photon kicks back the atom excited by
the first photon to the condensate, in addition to the process in which
two atoms are excited to the states $e^{i \Delta {\bf k}_1 \cdot {\bf R}}
f_0({\bf R})$ and $e^{i \Delta {\bf k}_2 \cdot {\bf R}} f_0({\bf R})$.
The Bose-enhancement factor of each process is
\begin{equation}
\langle N_0, 0_{\Delta {\bf k}_1} | \hat b_0^\dagger \hat b_{\Delta {\bf
k}_1} \hat b_{\Delta {\bf k}_1}^\dagger \hat b_0 | N_0, 0_{\Delta {\bf
k}_1} \rangle \simeq N
\end{equation}
and
\begin{eqnarray}
& & \langle (N - 2)_0, 1_{\Delta {\bf k}_1}, 1_{\Delta {\bf k}_2} | \hat
b_0^\dagger \hat b_{\Delta {\bf k}_1} \hat b_{\Delta {\bf k}_1}^\dagger
\hat b_0 | N_0, 0_{\Delta {\bf k}_1}, 0_{\Delta {\bf k}_2} \rangle
\nonumber \\
& & \simeq \sqrt{N(N - 1)},
\end{eqnarray}
and hence the probability is enhanced by a factor of two for large $N$.
Thus the above correlations are due to the quantum-statistical effect of
bosons.
In fact, the degree of second-order coherence reduces to $g^{(2)} = 1$
when $N = 1$, and a photon pair scattered from a single atom shows no
correlation.
We note that the presence of the condensate is required for the above
discussion.
The process in which the condensate participates is accompanied by the
Bose-enhancement factor of $\sqrt{N}$, and the probability that the
process occurs is $N$ times greater than that of the process between other
levels.
Therefore, scattering of photons by the condensate can clearly be
distinguished from that by the non-condensed atoms.
When the condensate is not present and the atoms are thermally
distributed, occupation numbers of any states are smaller than one, and
Bose-enhancement effects cannot be observed.

\begin{figure}[tb]
\begin{center}
\leavevmode\epsfxsize=86mm \epsfbox{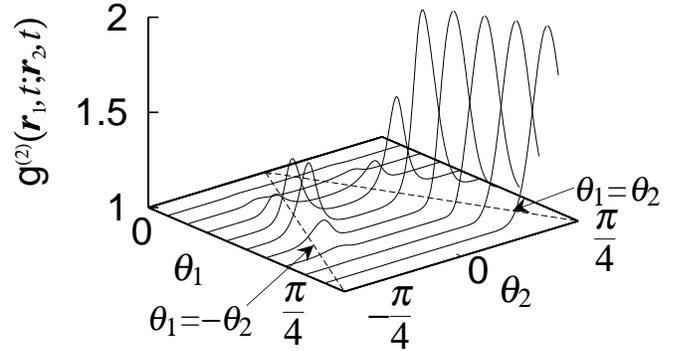}
\end{center}
\caption{
The angular dependence of the degree of second-order coherence
$g^{(2)}({\bf r}_1, t; {\bf r}_2, t)$ of photons scattered from $10^3$
BEC atoms.
The wavelength of the laser field $2 \pi / k_L = 589 {\rm nm}$ (the 3s-3p
transition of ${}^{23}{\rm Na}$) and  the characteristic length of the
trap $d = 1 \mu{\rm m}$ are assumed.
The wave vectors ${\bf k}_1$ and ${\bf k}_2$ are assumed to be in the same 
plane, and $\theta_1$ and $\theta_2$ denote their angles as measured from
${\bf k}_L$.
}
\label{f:g2}
\end{figure}

The degree of second-order coherence (\ref{g2_2}) for $t_1 = t_2$ is shown
in Fig.~\ref{f:g2}, where $N = 10^3$ and $d = 1 \mu{\rm m}$, and ${\bf
k}_1$ and ${\bf k}_2$ are assumed to be on the same plane.
We find that $g^{(2)} \simeq 1$ when the scattering angles are small, and
$g^{(2)}$ grows with increasing $\theta_1$.
While the peak of $g^{(2)}$ at the same scattering angle $\theta_1 =
\theta_2$ is $g^{(2)} \simeq 2$ for large $\theta_1$, the peak at the
opposite angle $\theta_1 = -\theta_2$ first grows and then decreases to 
$g^{(2)} \simeq 1$ as $\theta_1$ increases.
This is due to the fact that $\Delta {\bf k}_1 = -\Delta {\bf k}_2$ is not 
compatible with the energy conservation $|{\bf k}_1| = |{\bf k}_2| =
\omega_L / c$ (the recoil energy is much smaller than the photon energy),
and the third term in Eq.~(\ref{g2r}) cannot become unity even for
$\theta_1 = -\theta_2$.
In other words, the energy conservation prohibits the second photon from
exactly kicking the atom back to the condensate.

\begin{figure}
\begin{center}
\leavevmode\epsfxsize=86mm \epsfbox{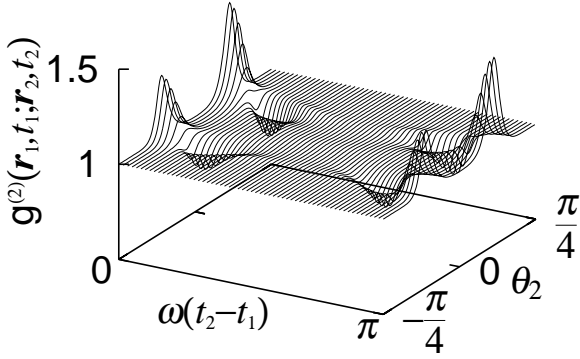}
\end{center}
\caption{
The time and angular dependence of the degree of second-order coherence
$g^{(2)}({\bf r}_1, t_1; {\bf r}_2, t_2)$ of photons scattered from $10^3$
BEC atoms.
The scattering angle of the first photon is fixed at $\theta_1 = \pi /
12$.
The wavelength of the laser field $2 \pi / k_L = 589 {\rm nm}$ (the 3s-3p
transition of ${}^{23}{\rm Na}$) and  the characteristic length of the
trap $d = 1 \mu{\rm m}$ are assumed.
The wave vectors ${\bf k}_1$ and ${\bf k}_2$ are assumed to be in the same 
plane, and $\theta_1$ and $\theta_2$ denote their angles as measured from
${\bf k}_L$.
}
\label{f:g2t}
\end{figure}

Figure \ref{f:g2t} shows the time dependence of the degree of second-order 
coherence (\ref{g2_2}).
The parameters are the same as those in Fig.~\ref{f:g2}, and the
scattering angle of the first photon is fixed at $\theta_1 = \pi / 12$.
We find that the second-order coherence is periodic in $t_2 - t_1$ with
the period given by $\pi \omega^{-1}$.
This periodic behavior reflects the fact that a recoil atom oscillates in
the trap.
The coherence disappears when the atom is kicked out of the condensate,
and revives when the atom returns to the condensate.
We note that $g^{(2)}$ goes to below unity before and after the appearance
of the peaks.
This indicates that photons are suppressed to be scattered into $\theta_2
\simeq \pm \theta_1$ in these time intervals.
This comes from the terms proportional to $N(N - 1)(N - 2)$ in the
numerator of Eq.~(\ref{g2_2}), which shows destructive interference of
probability amplitudes of two-photon processes in which one photon kicks
an atom and another photon does not disturb the atoms.

The degree of first-order coherence $g^{(1)}$ can be measured by
superposing the photon field in the directions of ${\bf k}_1$ and ${\bf
k}_2$ with a delay time $t_2 - t_1$ in the path of the ${\bf k}_1$
photon.
Taking a delay of order $\omega^{-1} \sim 1$ ms while retaining the
coherence might be feasible by reducing the speed of light using an
ultracold atomic gas~\cite{Hau}.
On the other hand, a measurement of $g^{(2)}$ does not require the
coherent delay, since it measures a correlation between detected photons.

\section{Effects of atom-atom interactions}
\label{s:interaction}

In this section, we take into account the atom-atom interaction on the
coherence properties of photons scattered from BEC within the Bogoliubov
approximation~\cite{Bogoliubov}.

The second-quantized form of the kinetic part of the Hamiltonian is given
by
\begin{eqnarray} \label{secondH}
\hat K & = & \int d{\bf R} \hat \psi^\dagger({\bf R}) \left[
-\frac{\hbar^2}{2m} \nabla^2 + V_t({\bf R}) - \mu \right] \hat \psi({\bf
R}) \nonumber \\
& & + \frac{U_0}{2} \int d{\bf R} \hat \psi^\dagger({\bf R}) \hat
\psi^\dagger({\bf R}) \hat \psi({\bf R}) \hat \psi({\bf R}),
\end{eqnarray}
where $\hat \psi({\bf R})$ is the field operator of the atoms (\ref{psi})
and $U_0 \equiv 4 \pi \hbar^2 a / m$, $a$ being the s-wave scattering
length.
In the Bogoliubov approximation, the atomic field operator is
divided into a mean field part and fluctuations from it as
\begin{equation} \label{psiB}
\hat \psi({\bf R}) = \sqrt{N} \psi_g({\bf R}) + \delta \hat \psi({\bf
R}),
\end{equation}
where $\psi_g({\bf R})$ is the normalized single-particle wave function
which obeys the Gross-Pitaevskii (GP) equation~\cite{GP}
\begin{equation} \label{GP}
\left[ -\frac{\hbar^2}{2m} \nabla^2 + V_t({\bf R}) + U_0 N |\psi({\bf
R})|^2 \right] \psi({\bf R}) = \mu \psi({\bf R}),
\end{equation}
where $\mu$ is determined so that the solution of Eq.~(\ref{GP}) satisfies
the normalization condition $\int d{\bf R} |\psi({\bf R})|^2 = 1$.
Equation (\ref{GP}) guarantees that if we substitute Eq.~(\ref{psiB}) into 
Eq.~(\ref{secondH}), terms linear in $\delta \hat\psi$ or
$\delta \hat\psi^\dagger$ in Eq.~(\ref{secondH}) vanish identically.
Because the interaction between atoms is weak, we may expect only small
deviations from the mean field at zero temperature, so that we may keep
only terms up to quadratic in $\delta \hat\psi$ and $\delta
\hat\psi^\dagger$:
\begin{eqnarray} \label{H2}
\hat K & \simeq &
N \int d{\bf R} \psi_g^*({\bf R}) \biggl[
-\frac{\hbar^2}{2m} \nabla^2 + V_t({\bf R}) - \mu \nonumber \\
& & + \frac{U_0 N}{2}
|\psi({\bf R})|^2 \biggr] \psi_g({\bf R}) \nonumber \\
& & + \int d{\bf R} \biggl[ \delta \hat\psi^\dagger({\bf R}) \biggl(
-\frac{\hbar^2}{2m} \nabla^2 + V_t({\bf R}) - \mu \nonumber \\
& & + 2 U_0 N |\psi({\bf
R})|^2 \biggr) \delta \hat\psi({\bf R}) \nonumber \\
& & + U_0 N \psi_g^{*2}({\bf R})
\delta \hat\psi^2({\bf R}) + U_0 N \psi_g^2({\bf R}) \delta
\hat\psi^{\dagger 2}({\bf R}) \biggr]
\end{eqnarray}
To diagonalize the quadratic terms, we write $\delta \hat \psi({\bf R})$
as
\begin{equation}
\delta \hat \psi({\bf R}) = \sum_\lambda \left[
u_\lambda({\bf R}) \hat \beta_\lambda + v_\lambda^*({\bf R}) \hat
\beta_\lambda^\dagger \right].
\end{equation}
Requiring the coefficients of the terms $\hat \beta_\lambda^2$ and $\hat
\beta_\lambda^{\dagger 2}$ to vanish, we find that $u_\lambda({\bf R})$
and $v_\lambda({\bf R})$ should satisfy
\begin{mathletters} \label{Beq}
\begin{eqnarray}
& & \left[ -\frac{\hbar^2}{2m} \nabla^2 + V_t({\bf R}) - \mu + 2 U_0 N
|\psi_g({\bf R})|^2 \right] u_\lambda({\bf R}) \nonumber \\
& & + U_0 N \psi_g({\bf R})^2
v_\lambda({\bf R}) = \hbar \omega_\lambda u_\lambda({\bf R}), \\
& & \left[ -\frac{\hbar^2}{2m} \nabla^2 + V_t({\bf R}) - \mu + 2 U_0 N
|\psi_g({\bf R})|^2 \right] v_\lambda({\bf R}) \nonumber \\
& & + U_0 N \psi_g^{*2}({\bf
R}) u_\lambda({\bf R}) = -\hbar \omega_\lambda v_\lambda({\bf R}).
\end{eqnarray}
\end{mathletters}
The quadratic part of the Hamiltonian (\ref{H2}) can then be diagonalized
as
\begin{equation} \label{HB}
\hat K = \sum_\lambda \hbar \omega_\lambda \hat \beta_\lambda^\dagger
\hat \beta_\lambda,
\end{equation}
where $\hat \beta_\lambda^\dagger$, $\hat \beta_\lambda$ are the creation
and annihilation operators of the Bogoliubov quasiparticles and $\hbar
\omega_\lambda$ denotes the energy of elementary excitations.
The constant terms are irrelevant to later discussions and are therefore
omitted in Eq.~(\ref{HB}).

The second quantized form of the operator $\sum_{j = 1}^N e^{i \Delta {\bf
k} \cdot {\bf R}_j(t)}$ that gives the momentum to an atom reads in the
Bogoliubov approximation as
\begin{equation} \label{DB}
\hat{\cal D}_B(\Delta {\bf k}, t) \equiv \int d{\bf R} \hat
\psi^\dagger({\bf R}, t) e^{i \Delta {\bf k} \cdot {\bf R}} \hat
\psi({\bf R}, t),
\end{equation}
where $\hat \psi({\bf R}, t) = e^{\frac{i}{\hbar} \hat K t} \hat
\psi({\bf R}) e^{-\frac{i}{\hbar} \hat K t}$.
Because $|u_\lambda({\bf R})| \gg |v_\lambda({\bf R})|$~\cite{Dalfovo97},
we may neglect $v_\lambda({\bf R})$ in Eq.~(\ref{psiB}).
The operator (\ref{DB}) then becomes
\begin{eqnarray} \label{DB2}
\hat{\cal D}_B(\Delta {\bf k}, t) & = & N \langle \psi_g | e^{i \Delta
{\bf k} \cdot {\bf R}} | \psi_g \rangle \nonumber \\
& & + \sqrt{N} \sum_\lambda \Bigl(
\langle \psi_g | e^{i \Delta {\bf k} \cdot {\bf R}} | u_\lambda \rangle
\hat \beta_\lambda e^{-i \omega_\lambda t} \nonumber \\
& & + \langle u_\lambda |
e^{i \Delta {\bf k} \cdot {\bf R}} | \psi_g \rangle \hat
\beta_\lambda^\dagger e^{i \omega_\lambda t} \Bigr) \nonumber \\ 
& & + \sum_{\lambda \mu} \langle u_\lambda | e^{i \Delta {\bf k} \cdot
{\bf R}} | u_\mu \rangle \hat \beta_\lambda^\dagger \hat \beta_\mu
e^{i (\omega_\lambda - \omega_\mu) t},
\end{eqnarray}
where we use the shorthand notation as in Eq.~(\ref{D}).
In the Bogoliubov approximation, the orthonormality condition is expressed
as
$\langle u_\lambda | u_\mu \rangle - \langle v_\lambda | v_\mu \rangle =
\delta_{\lambda \mu}$, and the completeness relation is given by
$\sum_\lambda (|u_\lambda \rangle \langle u_\lambda | - |v_\lambda \rangle
\langle v_\lambda |) = 1 - |\psi_g \rangle \langle \psi_g |$.
In the present approximation of neglecting $v_\lambda$, they reduce to
$\langle u_\lambda | u_\mu \rangle = \delta_{\lambda \mu}$ and
$\sum_\lambda |u_\lambda \rangle \langle u_\lambda | + |\psi_g \rangle
\langle \psi_g | = 1$, respectively.

Using the operator (\ref{DB2}), the source-field operator is expressed as
$\hat{\bf E}_{\rm s}^{(+)}({\bf r}, t) = {\bf F}({\bf r}, t) \hat{\cal
D}_B(\Delta {\bf k}, t)$.
The expectation values of the electric field and the intensity are
calculated to be
\begin{equation}
\langle \hat{\bf E}^{(+)}({\bf r}, t) \rangle =  {\bf F}({\bf r}, t) N
\langle \psi_g | e^{i \Delta {\bf k} \cdot {\bf R}} | \psi_g \rangle,
\end{equation}
\begin{equation}
I({\bf r}, t) = |{\bf F}({\bf r}, t)|^2 N \left( 1 + N |\langle \psi_g 
| e^{i \Delta {\bf k} \cdot {\bf R}} | \psi_g \rangle|^2 \right).
\end{equation}
These expressions are different from the noninteracting counterparts
(\ref{Eavg}) and (\ref{I}) in that the expectation value with
respect to the ground state of the noninteracting atoms $\langle 0
| e^{i \Delta {\bf k} \cdot {\bf R}} | 0 \rangle$ is replaced by that of
the ground state of the GP equation $\langle \psi_g  | e^{i \Delta {\bf k}
\cdot {\bf R}} | \psi_g \rangle$, and $N - 1$ by $N$.
The latter arises from the assumption of the Bogoliubov theory that
the ground state is a coherent state rather than a Fock state.
Although this is an artifact, the error is of order $1 / N$ and
negligible for $N \gg 1$.
The former reflects the fact that the ground-state wave function of
interacting atoms expands for the repulsive interaction $a > 0$, and
contracts for the attractive interaction $a < 0$, compared with that of
noninteracting atoms.

\begin{figure}
\begin{center}
\leavevmode\epsfxsize=86mm \epsfbox{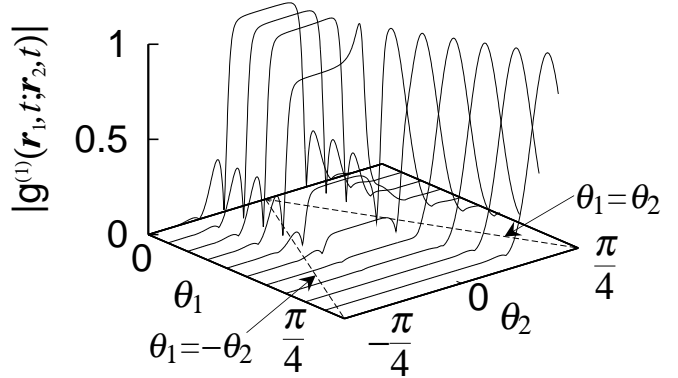}
\end{center}
\caption{
The angular dependence of the degree of first-order coherence
$g^{(1)}({\bf r}_1, t; {\bf r}_2, t)$ of photons scattered from $10^3$
interacting BEC atoms.
The s-wave scattering length is assumed to be $a = 2.75$ nm.
The wavelength of the laser field $2 \pi / k_L = 589 {\rm nm}$ (the 3s-3p
transition of ${}^{23}{\rm Na}$) and  the characteristic length of the
trap $d = 1 \mu{\rm m}$ are assumed.
The wave vectors ${\bf k}_1$ and ${\bf k}_2$ are assumed to be in the same 
plane, and $\theta_1$ and $\theta_2$ denote their angles as measured from
${\bf k}_L$.
}
\label{f:g1int}
\end{figure}

\begin{figure}
\begin{center}
\leavevmode\epsfxsize=86mm \epsfbox{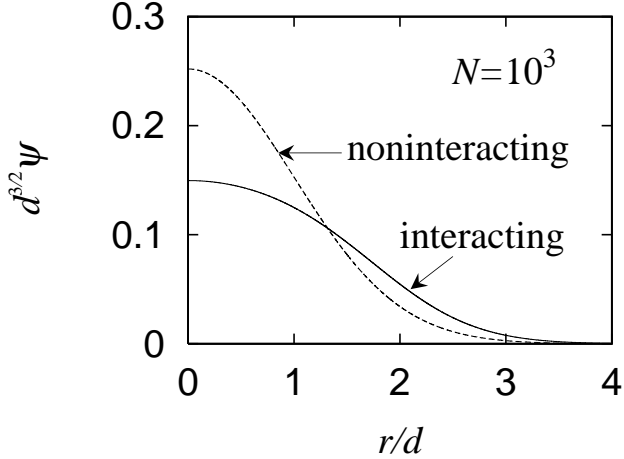}
\end{center}
\caption{
The profiles of the ground-state wave functions.
The solid curve shows the wave function of interacting $10^3$ atoms which
is obtained by solving the GP equation (\protect\ref{GP}),
and the dashed curve shows the noninteracting one.
}
\label{f:wf}
\end{figure}

The degree of first-order coherence is calculated to be
\begin{eqnarray} \label{g1int}
& & |g^{(1)}({\bf r}_1, t_1; {\bf r}_2, t_2)| = \nonumber \\
& & \frac{\langle e^{-i \Delta 
{\bf k}_1 \cdot {\bf R}} \hat U_B(t_1 - t_2) e^{i \Delta {\bf k}_2 \cdot
{\bf R}} \rangle_g + N \langle e^{-i \Delta {\bf k}_1 \cdot {\bf R}}
\rangle_g \langle e^{i \Delta {\bf k}_2 \cdot {\bf R}} \rangle_g}{\left( 1
+ N |\langle e^{i \Delta {\bf k}_1 \cdot {\bf R}} \rangle_g|^2
\right)^{\frac{1}{2}} \left( 1 + N |\langle e^{i \Delta {\bf k}_2 \cdot
{\bf R}} \rangle_g|^2 \right)^{\frac{1}{2}}}, \nonumber \\
\end{eqnarray}
where $\hat U_B(t) \equiv \sum_\lambda |u_\lambda \rangle e^{-i
\omega_\lambda t} \langle u_\lambda |$, and $\langle \cdots \rangle_g$
denotes the expectation value with respect to the ground state of the GP
equation (\ref{GP}).
To evaluate the expectation value that includes $\hat U_B$, we need to
find $u_\lambda$ which we obtain by numerically diagonalizing
Eqs.~(\ref{Beq}a) and (\ref{Beq}b) using the method discussed in
Ref.~\cite{Edwards}.
Figure~\ref{f:g1int} shows the degree of first-order coherence
(\ref{g1int}) for $t_1 = t_2$.
The parameters are taken to be $N = 10^3$, $d = 1 \mu{\rm m}$, and $a =
2.75$ nm.
The ground state of the GP equation (\ref{GP}) is obtained
numerically~\cite{Dalfovo96}.
The sharp dips in Fig.~\ref{f:g1int} correspond to the points at which the
sign of $g^{(1)}$ changes (note that Fig.~\ref{f:g1int} displays the
absolute value of $g^{(1)}$), while the curves in the noninteracting case
are smooth (see Fig.~\ref{f:g1}). 
To understand the oscillatory behaviors of $g^{(1)}$, we compare in
Fig.~\ref{f:wf} the profile of the interacting ground-state
wave function with $N = 10^3$ with the corresponding noninteracting one.
As we can see from the figure the density profile near the origin becomes
flat due to the repulsive interaction which gives rise to the oscillatory
behavior of $e^{i\Delta {\bf k} \cdot {\bf R}}$.
The validity of this interpretation can be confirmed analytically in the
limit of strong interaction $N a / d \gg 1$.
In this limit the kinetic energy may be neglected in comparison with the
trap potential and the interaction energy (Thomas-Fermi approximation),
and the wave function is given by
\begin{equation}
\psi_{\rm TF}(R) = \sqrt{\frac{15}{8 \pi R_0^5} (R_0^2 - R^2)},
\end{equation}
where $R_0 \equiv d (15 N a / d)^{1 / 5}$~\cite{Baym}.
In this Thomas-Fermi limit, the expectation value becomes
\begin{eqnarray} \label{TF}
\langle \psi_g | e^{i \Delta {\bf k} \cdot {\bf R}} | \psi_g \rangle & = &
\frac{15}{2 |R_0 \Delta {\bf k}|^5} \bigl[ (6 - 2 R_0 |\Delta {\bf k}|)
\sin R_0 |\Delta {\bf k}| \nonumber \\
& & - 6 R_0 |\Delta {\bf k}| \cos R_0 |\Delta {\bf
k}| \bigr],
\end{eqnarray}
which shows oscillations of $g^{(1)}$.
In contrast the Gaussian wave function gives a monotonically decreasing
nonzero value $\langle e^{i \Delta {\bf k} \cdot {\bf R}} \rangle = e^{-|d
\Delta {\bf k}|^2 / 4}$.

\begin{figure}
\begin{center}
\leavevmode\epsfxsize=86mm \epsfbox{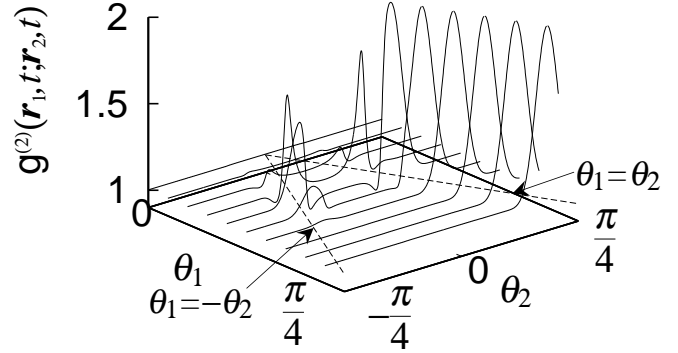}
\end{center}
\caption{
The angular dependence of the degree of second-order coherence
$g^{(2)}({\bf r}_1, t; {\bf r}_2, t)$ of photons scattered from $10^3$
interacting BEC atoms.
The s-wave scattering length is assumed to be $a = 2.75$ nm.
The wavelength of a laser field $2 \pi / k_L = 589 {\rm nm}$ (the 3s-3p
transition of ${}^{23}{\rm Na}$) and  the characteristic length of the
trap $d = 1 \mu{\rm m}$ are assumed.
The wave vectors ${\bf k}_1$ and ${\bf k}_2$ are assumed to be in the same 
plane, and $\theta_1$ and $\theta_2$ denote their angles as measured from
${\bf k}_L$.
}
\label{f:g2int}
\end{figure}

\begin{figure}
\begin{center}
\leavevmode\epsfxsize=86mm \epsfbox{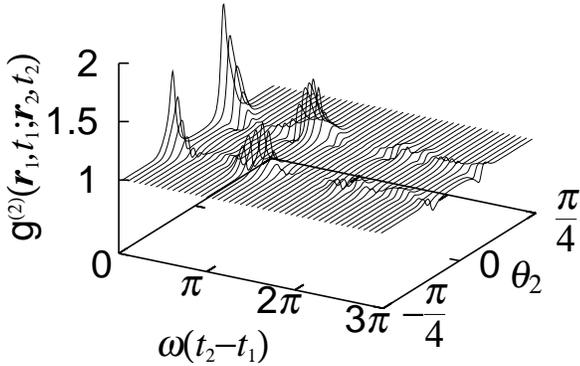}
\end{center}
\caption{
The time and angular dependence of the degree of second-order coherence
$g^{(2)}({\bf r}_1, t_1; {\bf r}_2, t_2)$ of photons scattered from $10^3$
interacting BEC atoms.
The scattering angle of the first photon is fixed at $\theta_1 = \pi /
12$.
The s-wave scattering length is assumed to be $a = 2.75$ nm.
The wavelength of the laser field $2 \pi / k_L = 589 {\rm nm}$ (the 3s-3p
transition of ${}^{23}{\rm Na}$) and  the characteristic length of the
trap $d = 1 \mu{\rm m}$ are assumed.
The wave vectors ${\bf k}_1$ and ${\bf k}_2$ are assumed to be in the same 
plane, and $\theta_1$ and $\theta_2$ denote their angles as measured from
${\bf k}_L$.
}
\label{f:g2tint}
\end{figure}

The degree of second-order coherence is calculated to be
\begin{eqnarray}
& & g^{(2)}({\bf r}_1, t_1; {\bf r}_2, t_2) = \nonumber \\
& & \Bigl[ N^4 |\langle e^{i
\Delta {\bf k}_1 \cdot {\bf R}} \rangle_g \langle e^{i \Delta {\bf k}_2
\cdot {\bf R}} \rangle_g|^2 \nonumber \\
& & + N^3 \Bigl( 2 {\rm Re} \langle e^{-i \Delta
{\bf k}_1 \cdot {\bf R}} \rangle_g \langle e^{-i \Delta {\bf k}_2 \cdot
{\bf R}} \rangle_g \nonumber \\
& & \times \langle e^{i \Delta {\bf k}_2 \cdot {\bf R}} \hat
U_B(t_2 - t_1) e^{i \Delta {\bf k}_1 \cdot {\bf R}} \rangle_g \nonumber \\ 
& & + 2 {\rm Re} \langle e^{-i \Delta {\bf k}_1 \cdot {\bf R}} \rangle_g
\langle e^{i \Delta {\bf k}_2 \cdot {\bf R}} \rangle_g \langle e^{-i
\Delta {\bf k}_2 \cdot {\bf R}} \hat U_B(t_2 - t_1) e^{i \Delta {\bf k}_1
\cdot {\bf R}} \rangle_g \nonumber \\
& & + |\langle e^{i \Delta {\bf k}_1 \cdot {\bf R}}
\rangle_g|^2 + |\langle e^{i \Delta {\bf k}_2 \cdot {\bf R}} \rangle_g|^2
\Bigr) \nonumber \\
\label{g2int}
& & + N^2 \Bigl( 1 + |\langle e^{i \Delta {\bf k}_2 \cdot {\bf R}} \hat
U_B(t_2 - t_1) e^{i \Delta {\bf k}_1 \cdot {\bf R}} \rangle_g|^2 \nonumber 
\\
& & + |\langle e^{-i \Delta {\bf k}_2 \cdot {\bf R}} \hat U_B(t_2 - t_1)
e^{i \Delta {\bf k}_1 \cdot {\bf R}} \rangle_g|^2 \nonumber \\
& & + 2 |\langle e^{i \Delta {\bf k}_1 \cdot {\bf R}} \rangle_g|^2 + 2
{\rm Re} \langle e^{-i \Delta {\bf k}_2 \cdot {\bf R}} \rangle_g \nonumber 
\\
& & \times \langle e^{-i \Delta {\bf k}_1 \cdot {\bf R}} \hat
U_B^\dagger(t_2 - t_1)
e^{i \Delta {\bf k}_2 \cdot {\bf R}} \hat U_B(t_2 - t_1) e^{i \Delta {\bf
k}_1 \cdot {\bf R}} \rangle_g \Bigr) \nonumber \\
& &  + N \Bigr] \Bigr/ \Bigl[ N^2 \left( 1 + N |\langle e^{i \Delta {\bf
k}_1 \cdot {\bf R}} \rangle_g|^2 \right) \left( 1 + N |\langle e^{i \Delta
{\bf k}_2 \cdot {\bf R}} \rangle_g|^2 \right) \Bigr]. \nonumber \\
\end{eqnarray}
Figure \ref{f:g2int} shows the degree of second-order coherence
(\ref{g2int}) for $t_1 = t_2$, where the parameters are the same as in
Fig.~\ref{f:g1int}.
One can see the complicated structure that arises from the oscillations
of $\langle e^{i \Delta {\bf k}_1 \cdot {\bf R}} \rangle_g$.
Figure~\ref{f:g2tint} shows the dependence on the time interval $t_2 -
t_1$ of $g^{(2)}({\bf r}_1, t_1; {\bf r}_2, t_2)$, where the parameters
are the same as in  Fig.~\ref{f:g1int}, and the scattering angle of the
first photon $\theta_1$ is fixed at $\pi / 12$.
We find that the peaks appear periodically with their heights
monotonically decreasing.
The decay of the peaks can be understood as follows.
The atom is kicked out of the condensate upon receipt of recoil momentum,
and oscillates in the trap potential while encountering the condensate
periodically.
If the atom-atom interaction is not present, the Gaussian wave function of 
the scattered atom retains its shape during the oscillations, and the
peaks revive completely as shown in Fig.~\ref{f:g2t}.
On the other hand, when the atom-atom interaction is present, the
wave function undergoes distortion by colliding the condensate.
As a result, the atom cannot be kicked back completely to the condensate
by scattering a photon, and therefore the correlation deteriorates.

\section{Conclusions}
\label{s:conclusion}

We have studied the coherence properties of photons scattered from the BEC
irradiated by a weak and far-off-resonant laser field, and have shown that
the degree of second-order coherence $g^{(2)}$ exhibits unique features
of trapped BEC.
That is, the probability of two photons being scattered either in the same
direction or in the symmetrical direction with respect to laser
propagation is enhanced, and these correlations appear periodically with
the period of atomic motion in the trapping potential.
These correlations in space and time are manifestations of the combined
effect of bosonic stimulation and quantized kinetic motion in the trap.
We have also taken into account effects of the atom-atom interaction, and
have confirmed that such correlations can be observed even in the presence 
of the atom-atom interaction with appearance of an additional structure
(sharp dips in Fig.~\ref{f:g1int}) resulting from the repulsive nature of 
the interaction.

\section*{ACKNOWLEDGMENT}

One of the authors (H.S.) acknowledges support by the Japan Society for
the Promotion of Science for Young Scientists.

\end{document}